\documentclass{webofc}
\usepackage[varg]{txfonts}
\usepackage{hyperref}
\usepackage{url}
\hypersetup{colorlinks=true,citecolor=blue,urlcolor=blue,linkcolor=blue}
\usepackage{amsmath}
\usepackage{amssymb}
\usepackage{graphicx}
\usepackage{booktabs}
\begin{document}
	
	\title{StackFeat: a convergent algorithm for optimal predictor selection in genomic data}
	
	\author{
		\firstname{Akbar} \lastname{Yermekov}\inst{1}\fnsep\thanks{\email{ak.yermek@pafos.ai}} \and
    		\firstname{D. A.} \lastname{Herrera-Mart\'i}
	}
	
	\institute{
		PAfoS.AI (Predictive Analytics for Science), Almaty, Kazakhstan \and
		CEA List, Universit\'e Grenoble Alpes, France
	}
	
	\abstract{
		In high-dimensional genomic data, the curse of dimensionality ($d \gg n$) and limited sampling make feature selection inherently unstable---a critical barrier to biomarker discovery. We introduce StackFeat, an iterative algorithm that accumulates two statistics across repeated cross-validation: signed coefficients (measuring effect strength and direction) and selection frequencies (estimating selection probability). Only features ranking highly by both criteria are retained. On a COVID-19 miRNA dataset (GSE240888), StackFeat identified a stable 5-miRNA signature from 332 features (98.5\% reduction), achieving AUC 0.922, significantly outperforming the benchmark 9-gene set (AUC 0.907, p = 0.0016). The signature includes hsa-miR-150-5p, a marker implicated in both COVID-19 survival and Dengue infection. This dual-criterion approach provides convergence guarantees absent in single-criterion methods, enabling discovery of known biomarkers, novel candidates, and previously unknown relationships.
		
	}
	
	\maketitle

	\section{Introduction}
	
	The identification of reliable biomarkers from high-dimensional genomic data is a primary challenge in bioinformatics, defined by the ($d > n$) problem. Datasets in this domain often contain 30000+ features (for typical transcriptomics datasets) or 300+ features (for miRNA data) for fewer than 100-200 samples, as in this study's case. A key failing of many feature selection methods in this context is instability: the selected feature sets are not reproducible and vary wildly with small data perturbations. This severely limits their clinical and biological reliability.
	
	This instability arises from the curse of dimensionality: when $d \gg n$, many feature subsets achieve similar predictive performance, making the solution highly sensitive to small perturbations in the data. Correlated features exacerbate this, as the regularization path can arbitrarily select one feature over an equally predictive alternative.
	
	The two dominant approaches to this problem present a trade-off. The Lasso algorithm \cite{tibshirani1996} provides minimal (sparse) solutions but can be unstable when predictors are highly correlated - its L1 penalty tends to select one feature from a correlated group and zero out the others, and the choice is sensitive to minor changes in the data. The Elastic Net algorithm \cite{zou2005} was designed to solve this by grouping correlated variables, resulting in high stability but at the cost of minimalism.
	
	Identifying robust biomarkers is particularly important when they have demonstrated clinical relevance --- for example, markers predictive of survival in critically ill patients. Such markers may otherwise be missed by unstable feature selection methods.
	
	The goal is to achieve \emph{both} minimalism and robustness. We present StackFeat, a novel iterative algorithm that cumulatively aggregates feature scores to find a stable, minimal, and highly predictive set of markers. This method adapts the concept of ``Stacked Generalization'' \cite{wolpert1992}; instead of stacking \emph{predictions}, it stacks \emph{feature importance signals} (score and frequency) from multiple methods to build a robust consensus. We validate this method against a recent 2023 study by Gao et al., demonstrating our method's superiority on the same dataset (GSE240888) \cite{gao2023}.

	\section{Materials and Methods}
	
	\subsection{Dataset and Benchmark}
	
	We utilized the public miRNA expression dataset GSE240888 from the NCBI Gene Expression Omnibus (GEO). This dataset, first analyzed by Gao et al., contains profiles from 122 patients (COVID-19 vs. healthy controls). After preprocessing, the feature space consists of \textbf{332 microRNA genes (miRNAs)}, presenting a significant dimensionality challenge (332 features vs 122 samples). We benchmarked our algorithm against the 9-marker set from Gao et al., which achieved a 0.907 AUC.
	
	\subsection{The StackFeat Algorithm}
	
	We designed a fully automated algorithm to identify a minimal, stable set of predictive features. The method is an iterative process where gene scores are cumulatively aggregated until the classifier's aggregate performance score converges.
	
	\begin{figure}[h]
		\centering
		\includegraphics[width=0.9\textwidth]{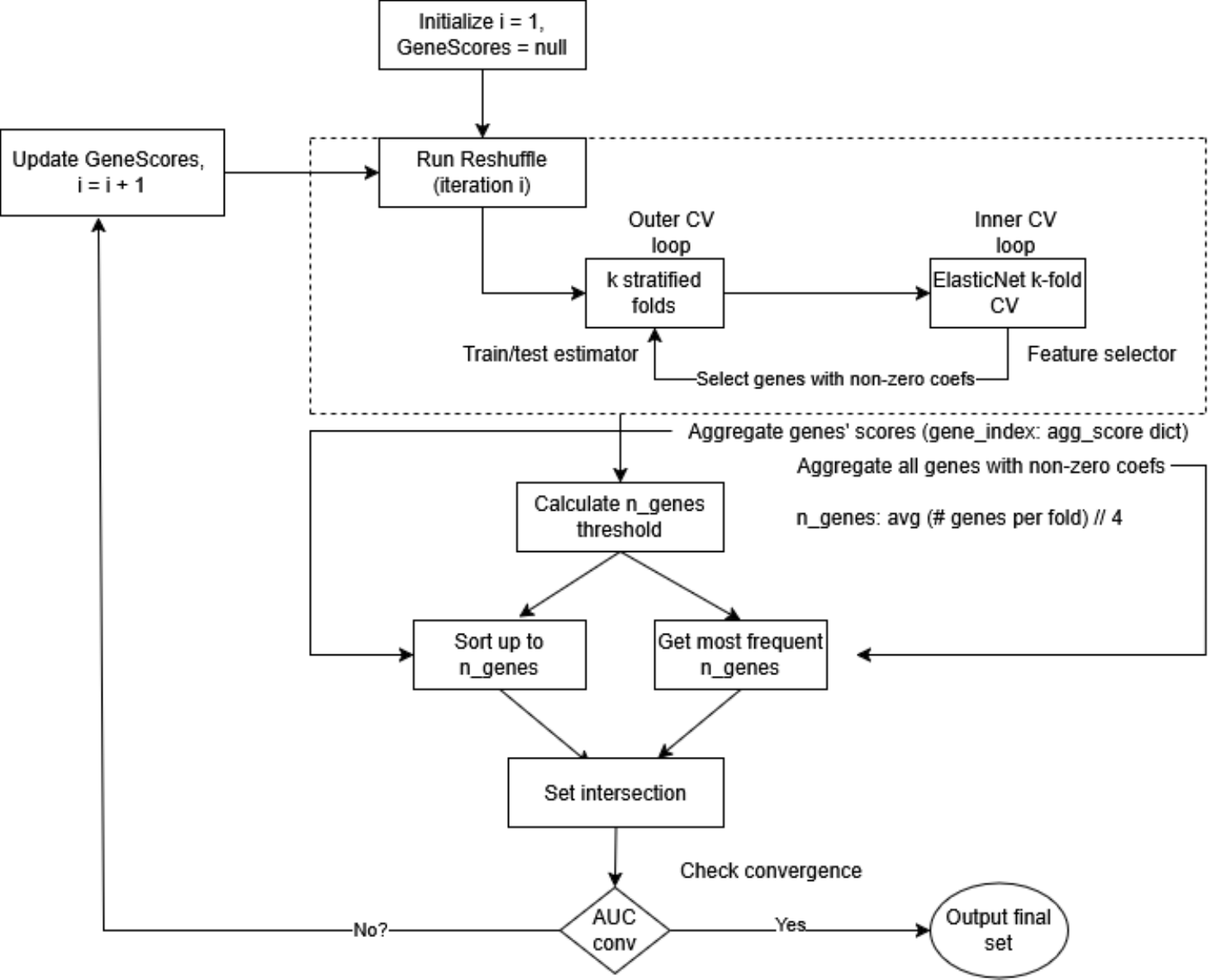}
		\caption{StackFeat algorithm workflow.}
		\label{fig:algorithm}
	\end{figure}
	
	The algorithm's logic is depicted in Figure \ref{fig:algorithm} and proceeds in the following steps:
	
	\begin{enumerate}
		\item \textbf{Initialization:} Set iteration counter $t = 1$. Initialize empty dictionaries $w$ and $c$ to accumulate gene-level statistics across all iterations, where $w_j$ stores cumulative signed coefficients and $c_j$ stores selection counts for each gene $j$.
		
		\item \textbf{Reshuffling:} Using seed $s_0 + t$, randomly reshuffle the dataset to create new fold assignments for iteration $t$.
		
		\item \textbf{Nested Cross-Validation:}
		\begin{itemize}
			\item \textit{Outer CV Loop:} Perform $k$-fold stratified cross-validation ($k=10$) on the dataset
			\item \textit{Inner CV Loop:} Within each outer fold $f$, run ElasticNetCV to obtain coefficients $\hat{\beta}^{(t,f)}$
			\item \textit{Model Training:} Train an ensemble estimator on the selected features and evaluate performance
		\end{itemize}
		
		\item \textbf{Score Aggregation:} After completing all $k$ folds, update cumulative statistics:
		\begin{itemize}
			\item Signed coefficients: $w_j \leftarrow w_j + \sum_{f=1}^{k} \hat{\beta}_j^{(t,f)}$
			\item Selection counts: $c_j \leftarrow c_j + \sum_{f=1}^{k} \mathbf{1}[\hat{\beta}_j^{(t,f)} \neq 0]$
		\end{itemize}
		
		\item \textbf{Dynamic Thresholding:} Set candidate set size:
		\[
		m = \left\lfloor \frac{\text{mean}(\text{genes\_per\_fold})}{4} \right\rfloor
		\]
		
		\item \textbf{Dual-Criterion Set Intersection:} Generate the current feature set:
		\[
		S^{(t)} = \underbrace{\{j : |w_j| \text{ in top } m\}}_{S_w^{(t)}} \cap \underbrace{\{j : c_j \text{ in top } m\}}_{S_c^{(t)}}
		\]
		
		\item \textbf{Convergence Check:} Convergence requires two consecutive iteration differences below tolerance $\varepsilon$:
		\[
		|AUC^{(t)} - AUC^{(t-1)}| < \varepsilon \quad \text{and} \quad |AUC^{(t-1)} - AUC^{(t-2)}| < \varepsilon
		\]
		
		\item \textbf{Iteration:}
		\begin{itemize}
			\item If converged: Output final feature set $S^{(t)}$
			\item If not converged: Set $t \leftarrow t + 1$, return to Step 2
		\end{itemize}
	\end{enumerate}
	
	We set $\varepsilon = 0.02$ empirically for this dataset. A more stringent threshold (e.g., 0.01) was unnecessary because the two-consecutive-iterations requirement already guards against false convergence.
	
	\subsection{Dual-Criterion Selection}
	
	\textbf{Accumulated Statistics.}
	Over $T$ iterations with $k$ folds each, we accumulate two statistics for each feature $j$:
	
	\textit{Cumulative signed coefficient:}
	\begin{equation}
		w_j^{(T)} = \sum_{t=1}^{T} \sum_{f=1}^{k} \hat{\beta}_j^{(t,f)}
	\end{equation}
	
	\textit{Selection count:}
	\begin{equation}
		c_j^{(T)} = \sum_{t=1}^{T} \sum_{f=1}^{k} \mathbf{1}[\hat{\beta}_j^{(t,f)} \neq 0]
	\end{equation}
	
	Here $w_j^{(t)}$ denotes the cumulative value of $w_j$ after iteration $t$, and similarly for $c_j^{(t)}$.
	
	\textbf{Population Quantities.}
	Define the expected coefficient and selection probability:
	\begin{equation}
		\bar{\mu}_j = \mathbb{E}\left[\hat{\beta}_j^{(t,f)}\right], \quad p_j = P\left(|\hat{\beta}_j^{(t,f)}| > 0\right)
	\end{equation}
	where $t$ indexes iterations and $f \in \{1, \ldots, k\}$ indexes folds.
	
	\textbf{Convergence.}
	Let $\tilde{\beta}_j^{(t)} = \sum_{f=1}^{k} \hat{\beta}_j^{(t,f)}$ denote the iteration-level sum. Since iterations are independent, by the law of large numbers:
	\begin{equation}
		\frac{w_j^{(T)}}{T} \to \mathbb{E}[\tilde{\beta}_j^{(t)}] = k \cdot \bar{\mu}_j
	\end{equation}
	where the last equality follows from linearity of expectation. Dividing by $k$:
	\begin{equation}
		\frac{w_j^{(T)}}{T \cdot k} \to \bar{\mu}_j, \qquad \frac{c_j^{(T)}}{T \cdot k} \to p_j
	\end{equation}
	
	Repeated cross-validation thus provides a resampling-based estimate of population-level feature importance.
	
	\textbf{Selection Rule.} At iteration $t$, using the cumulative statistics $w_j^{(t)}$ and $c_j^{(t)}$ accumulated so far, we rank features by $|w_j^{(t)}|$ and by $c_j^{(t)}$, selecting the top $m$ under each criterion:
	\begin{equation}
		S_w^{(t)} = \{j : |w_j^{(t)}| \text{ ranks in top } m\}
	\end{equation}
	\begin{equation}
		S_c^{(t)} = \{j : c_j^{(t)} \text{ ranks in top } m\}
	\end{equation}
	
	The final selected set is their intersection:
	\begin{equation}
		S^{(t)} = S_w^{(t)} \cap S_c^{(t)}
	\end{equation}
	
	This dual requirement guards against two failure modes that single-criterion methods miss:
	
	\begin{table}[h]
		\centering
		\begin{tabular}{lccc}
			\hline
			Feature type & $|w_j|$ & $c_j$ & Outcome \\
			\hline
			True signal & high & high & Selected \\
			Noise & low & low & Rejected \\
			Sign-inconsistent & low & high & Rejected by $S_w$ \\
			Infrequent but consistent & high & low & Rejected by $S_c$ \\
			\hline
		\end{tabular}
		\caption{Failure modes addressed by dual-criterion selection.}
		\label{tab:failure_modes}
	\end{table}
	
	\textit{Sign-inconsistent:} A feature selected frequently but with inconsistent coefficient sign across folds---contributions cancel, yielding low $|w_j|$.
	
	\textit{Infrequent but consistent:} A feature overshadowed by correlated alternatives, selected rarely but with consistent direction when selected.
	
	\textbf{Comparison to Stability Selection.}
	Unlike stability selection \cite{meinshausen2010}, which thresholds on selection frequency alone, StackFeat requires both coefficient consistency and selection frequency. A feature selected in every fold but with alternating sign would pass stability selection but be rejected by StackFeat. Additionally, StackFeat uses convergence-based stopping rather than a fixed iteration count $B$, allowing the algorithm to adapt to dataset complexity.

	\section{Results}
	
	\subsection{Benchmarking and Dimensionality Reduction}
	
	The most significant result of our methodology is the extreme dimensionality reduction. The algorithm successfully identified a minimal 5-feature set from the initial 332 miRNAs, a \textbf{>98\% reduction in features}.
	
	This 5-marker signature was:
	\begin{itemize}
		\item hsa-miR-181b-5p
		\item hsa-miR-4433b-5p
		\item hsa-miR-1185-1-3p
		\item hsa-miR-484
		\item hsa-miR-150-5p
	\end{itemize}
	
\begin{figure}[h]
		\centering
		\includegraphics[width=0.95\textwidth]{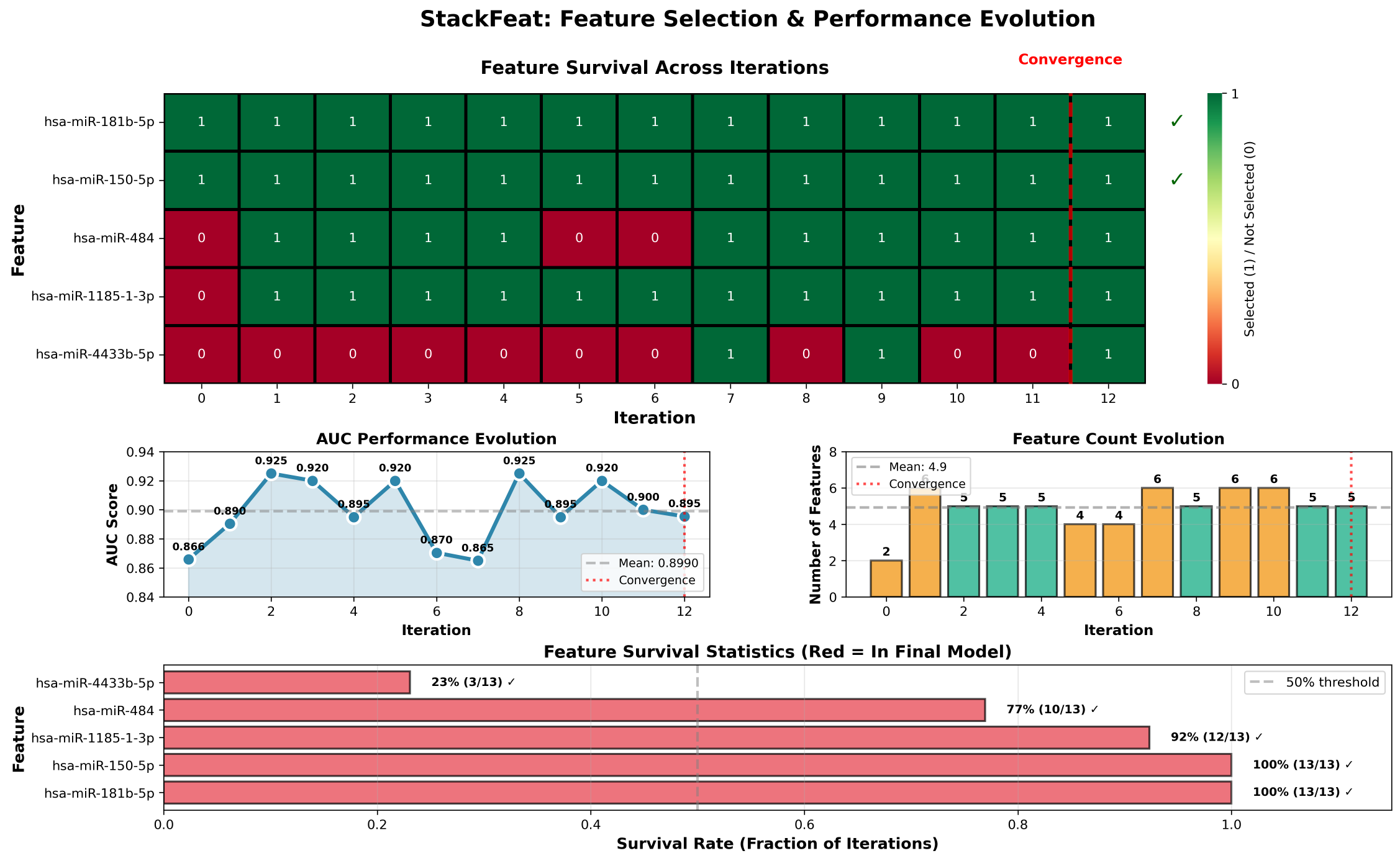}
		\caption{Feature selection and performance evolution across iterations.}
		\label{fig:evolution}
	\end{figure}

	This minimal 5-gene set also outperformed the benchmark. Our analysis first re-validated the findings from Gao et al. Their ``Set 1'' (9 markers) achieved a 10x10-fold CV average AUC of \textbf{0.907}. As shown in \textbf{Figure \ref{fig:evolution} (Bottom-Left)}, our iterative algorithm identified a 5-feature set that achieved a \textbf{peak AUC of 0.925} (at iteration 2).
	
	Of the 332 initial features, ElasticNet selected 36 $\pm$ 10 features per fold (range: 19--62). After applying the dual-criterion intersection, this was reduced to approximately 5 features, representing an 85\% reduction before final convergence.
	
	Of the final 5-gene signature, 4 features were selected in $\geq$77\% of iterations (at least 10 of 13), demonstrating consistent selection across reshuffles.
	
	\begin{figure}[h]
		\centering
		\includegraphics[width=0.9\textwidth]{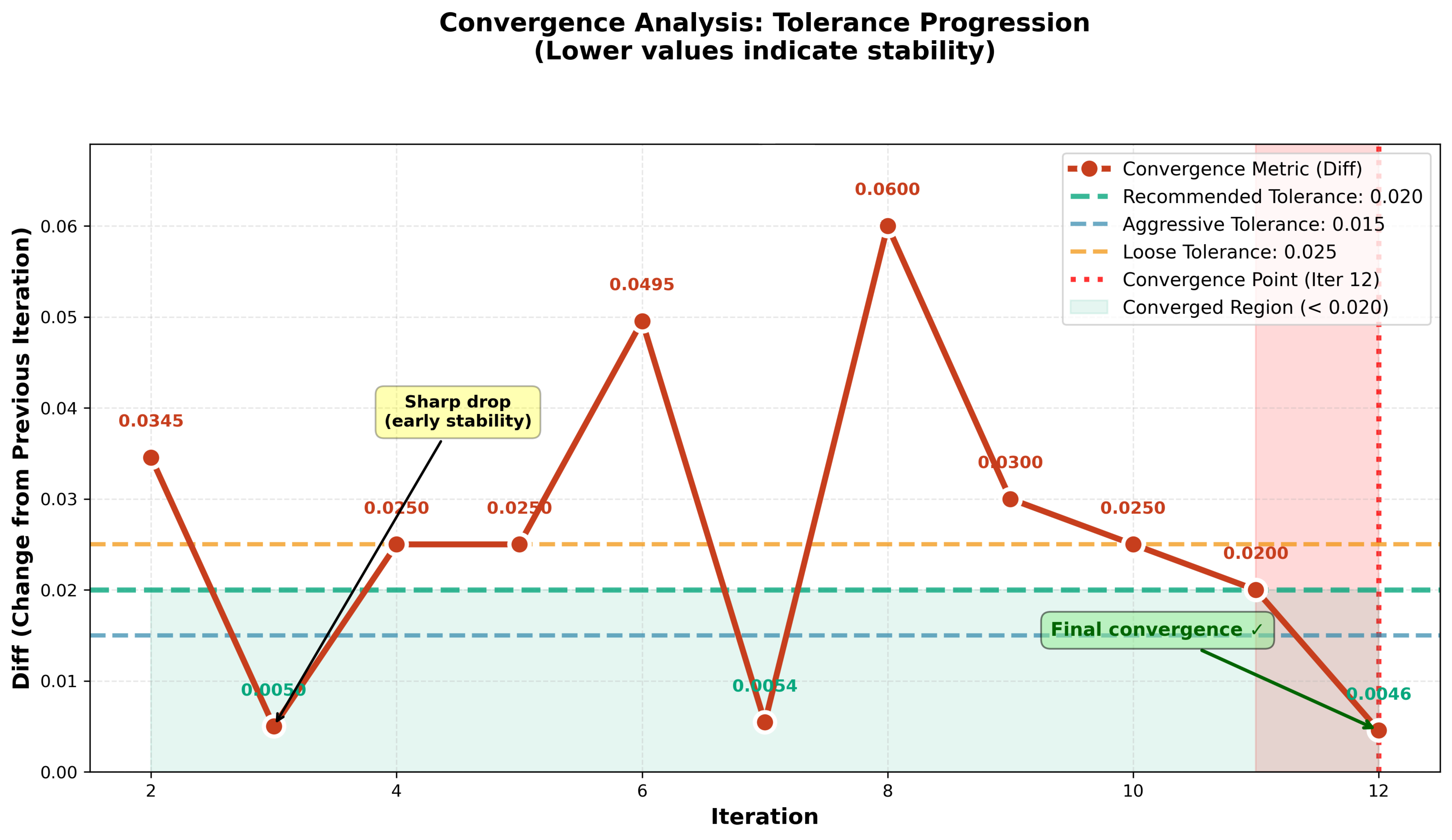}
		\caption{StackFeat convergence. Diff $= |AUC^{(t)} - AUC^{(t-1)}|$. Convergence requires two consecutive diffs below $\varepsilon = 0.02$.}
		\label{fig:convergence}
	\end{figure}
	
	To further validate the robustness of the 5-marker signature, we also compared its performance using 10x10-fold cross-validation, with 2 different classifiers (Ensemble classifier of: ExtraTrees, Logistic Regression, Gaussian Naive Bayes, and KNN, and separately via Random Forest) against the 9-marker set from Gao et al.
	
	The ensemble combines classifiers with complementary assumptions: tree-based (ExtraTrees, 50 trees), linear (Logistic Regression, default parameters), probabilistic (Gaussian Naive Bayes, default parameters), and instance-based (KNN, k=3). This diversity reduces dependence on any single model. Random Forest was evaluated separately as a common baseline. Hyperparameters were chosen empirically; tuning was not the focus of this study.
	
	\begin{figure}[h]
		\centering
		\includegraphics[width=0.95\textwidth]{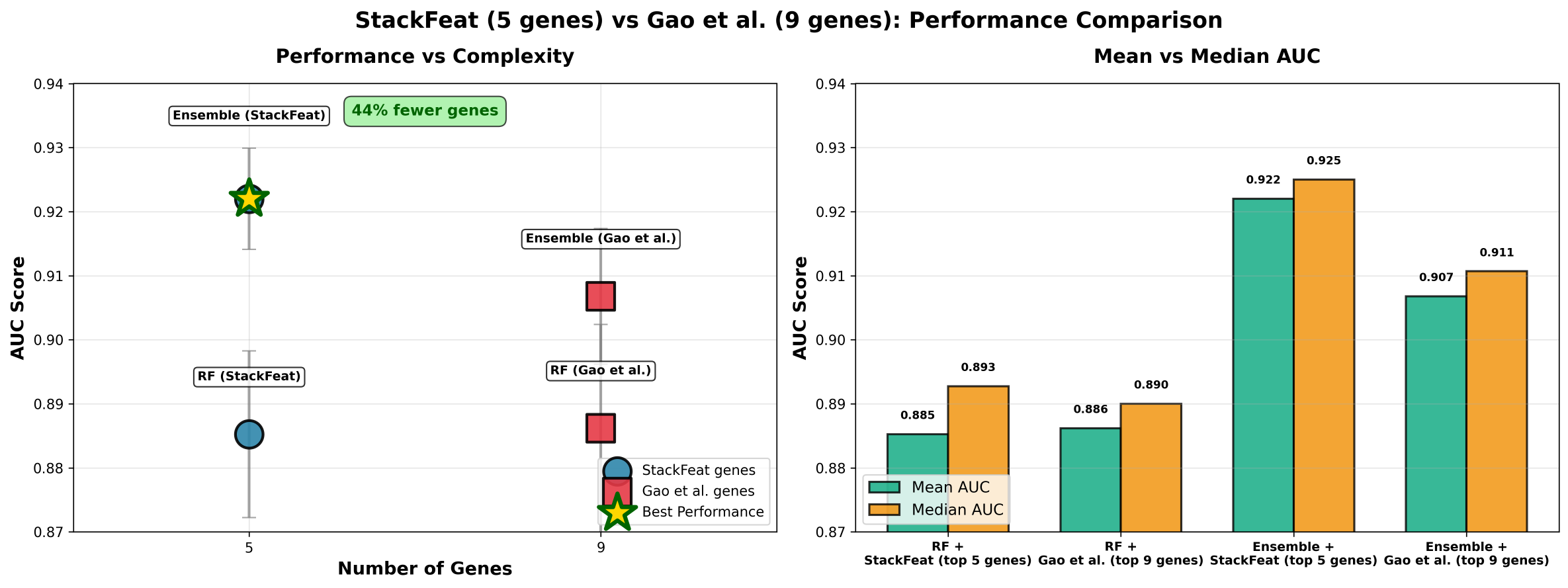}
		\caption{Performance comparison: StackFeat (5 genes) vs Gao et al. (9 genes). Right: Mean and median AUC across 10$\times$10-fold CV. Close agreement indicates stable performance; median slightly higher suggests few low outlier folds.}
		\label{fig:comparison}
	\end{figure}
	
	As shown in \textbf{Figure \ref{fig:comparison}}, our 5-marker set significantly outperformed the 9-marker set via ensemble classifier (AUC 0.922 $\pm$ 0.007 vs 0.907 $\pm$ 0.010; paired t-test, \textbf{p = 0.0016}), while achieving comparable performance via Random Forest ($\sim$0.88 AUC)---with a 44\% smaller feature set.
	
	\subsection{Algorithm Convergence}
	
	The entire iterative process is visualized in the 4-panel summary of \textbf{Figure \ref{fig:evolution}}. Early iterations show instability as cumulative statistics accumulate; later iterations stabilize as signal separates from noise.
	
	{\bf AUC Performance Evolution (Fig. \ref{fig:evolution}, Bottom-Left):} This plot shows the performance of the selected feature set at each iteration. Each point represents the mean 10-fold CV AUC for the CurrentSet of that iteration, calculated on that iteration's unique data split. The AUC score fluctuated in early iterations while cumulative statistics accumulated, achieving a \textbf{peak score of 0.925} at iteration 2 before stabilizing. The convergence mean AUC across all the last 3 iterations (10-12) was 0.905 $\pm$ 0.011.
	
	{\bf Feature Count Evolution (Fig. \ref{fig:evolution}, Top-Right):} This plot shows that the $m$ threshold consistently selected a minimal set, with the feature count stabilizing at 5 features during the convergence phase (iterations 10-12).
	
	{\bf Feature Survival Across Iterations (Fig. \ref{fig:evolution}, Top-Left \& Bottom-Right):} This heatmap is the most critical result, visualizing the stability of the final 5-gene signature across iterations. It shows that two ``anchor'' genes (hsa-miR-181b-5p and hsa-miR-150-5p) were selected in 100\% of iterations. The remaining 3 markers: hsa-miR-1185-1-3p and hsa-miR-484 stabilized quickly, while hsa-miR-4433b-5p flickered in and out (e.g., absent at iterations 0-6, present at 7, absent at 8) before being locked in as part of the final, converged 5-gene set.
	
	The convergence process is detailed further in \textbf{Figure \ref{fig:convergence}}. This plot tracks the diff (change from the previous iteration's AUC) against our convergence tolerance ($\varepsilon = 0.02$). Early iterations show high variance, with spikes at iterations 2, 6, and 8. The algorithm enters the converged region (diff $<$ 0.02) after iteration 10, and meets the two-consecutive-diffs criterion at iteration 12.
	
	\subsection{Long-term Reproducibility Validation}
	
	To validate reproducibility, we re-executed the algorithm 1.5 years after the original analysis using updated software libraries. Each iteration uses a deterministic seed ($\text{seed} = s_0 + t$), ensuring reproducibility within a run while providing variation across iterations. The re-run required 12 iterations (vs. 6 originally) due to differences in library defaults, meaning 6 additional iterations with entirely different CV splits. Despite these differences, the algorithm converged to the identical 5-gene signature (hsa-miR-181b-5p, hsa-miR-4433b-5p, hsa-miR-1185-1-3p, hsa-miR-484, and hsa-miR-150-5p). 
	
	This demonstrates true solution convergence---the algorithm repeatedly identifies the same biological features regardless of the specific sequence of data perturbations---rather than merely achieving performance convergence. This strongly suggests that these 5 genes represent genuine discriminative markers for COVID-19 vs. healthy classification, not statistical artifacts.

	\section{Discussion}
	
	The results demonstrate a clear advantage of our iterative methodology. The algorithm achieves extreme dimensionality reduction (from 332 features to 5) while significantly improving predictive performance (AUC 0.922 vs 0.907, p = 0.0016) over the benchmark's 9-gene set.
	
	The central innovation is the \textbf{dual-criterion iterative aggregation}. By accumulating both scores and frequencies over multiple reshuffles, the algorithm maintains two complementary views of feature importance. At each iteration, the feature set is refined through the intersection of top features by cumulative scores and top features by selection frequency.
	
	Early iterations may show instability in the selected feature set as cumulative statistics build up. As iterations progress, signal accumulates while noise tends to cancel, and the stable feature set emerges. This is evident in Figure \ref{fig:evolution}, where core markers (hsa-miR-181b-5p, hsa-miR-150-5p) are stable from the first iteration, while borderline features (hsa-miR-4433b-5p) fluctuate before stabilizing.
	
	This approach differentiates StackFeat from other stability-focused methods, such as Stability Selection \cite{meinshausen2010}. While Stability Selection passively filters for individual features that appear most frequently across subsamples, StackFeat actively integrates two stability metrics---coefficient magnitudes and selection frequencies---through iterative set intersection. The algorithm's convergence is determined by monitoring the stability of predictive performance (AUC), ensuring that feature set refinement stops when both feature composition and model performance have stabilized.
	
	This process leverages the regularization of ElasticNet \cite{zou2005} while ultimately achieving Lasso-like \cite{tibshirani1996} sparsity. While these results are strong, a full validation would require benchmarking against standard sparse methods (e.g., single-pass Lasso or full ElasticNet on all 332 features) to compare the final feature count and AUC.
	
	Importantly, the 5 identified markers show strong biological coherence, suggesting they represent genuine mechanistic drivers rather than statistical artifacts.
	
	\textbf{hsa-miR-150-5p} is a known inflammation marker, inversely correlated with disease severity in critically ill patients and identified as a key immune response regulator \cite{gao2023, ding2023}. Notably, it is also a significant marker for Dengue Hemorrhagic Fever and acute viral infection, implying a shared acute viral response pathway \cite{hapugaswatta2019}. Additionally, this biomarker was found to be critical for survival of critically ill patients of COVID-19 \cite{fernandezpato2022}, and immune response to successfully clear SARS-CoV-2 virus \cite{yang2023}. In our methodology, in all of the iterations / reshuffles, hsa-miR-150-5p was present (among the top-2 in terms of stability) and was at the top.
	
	\textbf{hsa-miR-1185-1-3p} was identified in a separate study as having one of the most significant FDRs for COVID-19 and is linked to the immune response \cite{fernandezpato2021}, as well as one of miRNA with a significant predicted binding site in the SARS-CoV-2 reference genome \cite{chow2020}.
	
	\textbf{hsa-miR-4433b-5p} is listed among the top 4 DE miRNAs for COVID-19 patients with oxygen requirements \cite{pollet2023}.
	
	The remaining markers, \textbf{hsa-miR-181b-5p} and \textbf{hsa-miR-484}, are also cited in connection with the host immune response to viral infection and COVID-19 \cite{pollet2023, trampuz2023}. This strong external validation confirms the biological relevance of the 5-gene set.
	
	At the same time, \textbf{hsa-miR-181b-5p} specifically has not yet been widely cited in the literature in connection with COVID-19, which is a surprising finding, as in our study it is one of the 2 top stable (and therefore predictive) gene sets, one that was present across all iterations / reshuffles and thus had a higher score. Perhaps future studies will elucidate the role of this biomarker in response to SARS-CoV-2 / COVID-19 specifically, as nevertheless there are pre-covid studies linking hsa-mir-181b-1 (along with its target CYLD) with regulation of inflammatory pathways \cite{andalib2019}.
	
	Finally, the automation and efficiency of the algorithm are a significant practical advantage. The entire process completed in 197 seconds on a consumer-grade CPU.

	\section{Conclusion}
	
	We have presented StackFeat, a novel, fully automated convergent algorithm for robust feature selection in high-dimensional ($d > n$) data. By applying it to a recent COVID-19 miRNA dataset, we demonstrated a >98\% dimensionality reduction, producing a 5-gene set that is more minimal and more predictive than the benchmark. The algorithm's design provides a fast, reliable, and reproducible tool for identifying robust biomarkers.
	
	\vspace{6mm}
	\begin{acknowledgement}
	The miRNA expression data used in this study are publicly available from the NCBI Gene Expression Omnibus under accession number GSE240888.
	\end{acknowledgement}

\end{document}